\documentclass[12pt,a4paper]{article}
\usepackage{arxiv}

\usepackage[utf8]{inputenc}
\usepackage[english]{babel}
\usepackage{amsmath,amssymb}
\usepackage{graphicx}
\usepackage{setspace}
\usepackage{hyperref}
\usepackage{natbib}
\usepackage{booktabs}
\usepackage{caption}

\hypersetup{
    colorlinks=true,
    linkcolor=blue,
    citecolor=blue,
    urlcolor=blue
}

\title{\textbf{Hardware-Level Governance of AI Compute:\\A Feasibility Taxonomy for Regulatory Compliance and Treaty Verification}}

\author{
	Samar Ansari\\
	School of Computing and Engineering Sciences\\
	University of Chester\\
	Chester, CH1 4BJ, United Kingdom \\
	\texttt{m.ansari@chester.ac.uk}
}

\begin{document}

\maketitle


\begin{abstract}
The governance of frontier AI increasingly relies on controlling access to computational resources, yet the hardware-level mechanisms invoked by policy proposals remain largely unexamined from an engineering perspective. This paper bridges the gap between AI governance and computer engineering by proposing a taxonomy of 20 hardware-level governance mechanisms, organised by function (monitoring, verification, enforcement) and assessed for technical feasibility on a four-point scale from currently deployable to speculative. For each mechanism, we provide a technical description, a feasibility rating, and an identification of adversarial vulnerabilities. We map the taxonomy onto four governance scenarios: domestic regulation, bilateral agreements, multilateral treaty verification, and industry self-regulation. Our analysis reveals a structural mismatch: the mechanisms most needed for treaty verification, including on-chip compute metering, cryptographic proof-of-training, and hardware-embedded enforcement, are also the least mature. We assess principal threats to compute-based governance, including algorithmic efficiency gains, distributed training methods, and sovereignty concerns. We identify a temporal constraint: the window during which semiconductor manufacturing concentration makes hardware-level governance implementable is narrowing, while R\&D timelines for critical mechanisms span years. We present an adversary-tiered threat analysis distinguishing commercial, non-state, and nation-state actors, arguing the appropriate security standard is tamper-evident assurance analogous to IAEA verification rather than absolute tamper-proofing. The taxonomy, feasibility classification, and mechanism-to-scenario mapping provide a technical foundation for policymakers and identify the R\&D investments required before hardware-level governance can support verifiable international agreements.
\end{abstract}



\section{Introduction}
\label{sec:introduction}

The governance of frontier artificial intelligence systems increasingly depends on controlling access to the computational resources required to develop them. Compute, the aggregate processing power used to train AI models, has emerged as the most prominent regulatory lever in both enacted legislation and proposed international agreements. The EU AI Act imposes obligations on general-purpose AI models trained above $10^{25}$~floating-point operations~\cite{euaiact2024}. US Executive Order~14110 established reporting requirements tied to compute thresholds and cloud infrastructure capacity~\cite{whitehouse2023eo}. Proposals for international AI governance, including analogues to the IAEA and the Nuclear Non-Proliferation Treaty, increasingly feature hardware-level verification as a key component of their compliance architectures~\cite{belfield2025domestic, ramiah2025toward, baker2023nuclear}.

The rationale for compute-based governance rests on three properties that distinguish compute from other potential governance targets such as training data, algorithms, or trained models: compute is \textit{detectable} (training runs require physical infrastructure that consumes large amounts of power), \textit{excludable} (access to advanced chips can be restricted through export controls and licensing), and \textit{quantifiable} (floating-point operations provide a measurable proxy for AI capability)~\cite{sastry2024computing, heim2024govern}. These properties make compute governance what Heim et al.\ describe as a ``particularly important'' approach to AI governance, feasible in ways that alternatives based on data, algorithms, or trained models are not~\cite{heim2024govern}.

However, the governance literature that discusses hardware-level mechanisms is almost entirely written by policy researchers, legal scholars, and political scientists who do not engage with the technical substrate in detail. The result is a body of policy proposals that reference ``on-chip metering,'' ``cryptographic attestation,'' ``hardware kill switches,'' and ``trusted execution environments'' without assessing whether these mechanisms can actually be built, at what cost, on what timeline, and against what threat model. Conversely, the computer engineering literature on hardware security, trusted execution environments, and semiconductor supply chains rarely engages with governance applications. This paper bridges that gap.

We propose a taxonomy of 20~hardware-level governance mechanisms, organised by governance function (monitoring, verification, enforcement) and assessed for technical feasibility on a four-point scale from currently deployable to speculative. For each mechanism, we provide a technical description grounded in the engineering literature, an evidence-based feasibility rating, an identification of the governance scenarios it enables, and an honest assessment of its limitations and adversarial vulnerabilities. We then map this taxonomy onto four governance scenarios (domestic regulation, bilateral agreements, multilateral treaty verification, and industry self-regulation) to assess which mechanisms are appropriate and feasible for each context.

The core contributions are fourfold. First, we provide a \textit{systematic feasibility classification} that distinguishes between mechanisms at very different stages of readiness, a distinction that existing literature frequently obscures by describing all proposals at a uniform level of abstraction. Second, we present an \textit{adversary-tiered threat analysis} that explicitly distinguishes commercial actors, well-resourced non-state actors, and nation-state adversaries, arguing that the appropriate security standard for governance hardware is not absolute tamper-proofing but tamper-evident assurance analogous to IAEA verification. Third, we construct a \textit{mechanism-to-scenario mapping} that provides direct policy utility by identifying which technical investments are necessary for which governance objectives. Fourth, and perhaps most distinctively, we identify a \textit{temporal constraint}: the mechanisms most needed for international treaty verification (on-chip compute metering, cryptographic proof-of-training, hardware-embedded enforcement) are also the least mature, and their development timelines of 18~months to 4~years of R\&D followed by a further 4~years for sufficiently widespread deployment~\cite{aarne2024secure} are on the same order as the window during which semiconductor manufacturing concentration makes hardware-level governance implementable. This convergence of timelines defines a critical and time-bounded R\&D agenda.

\paragraph{Scope and method.} The taxonomy was constructed through a systematic review of the compute governance literature (2022--2026) and the relevant computer engineering literature on hardware security, trusted execution environments, and semiconductor supply chains. Mechanisms were identified by surveying all hardware-level governance proposals in the reviewed literature and consolidating overlapping proposals into distinct mechanism categories. Feasibility ratings were assigned using a four-point scale adapted from technology readiness level (TRL) frameworks: \textit{currently deployable} (TRL~7--9; functional prototypes or production systems exist), \textit{near-term} (TRL~4--6; requires engineering effort but no fundamental research), \textit{requires R\&D} (TRL~2--3; technically plausible but lacks working prototypes), and \textit{speculative} (TRL~1; conceptual proposals without clear implementation pathways). Ratings reflect the most advanced publicly documented state of each mechanism as reported in the reviewed literature; where sources disagreed on readiness, we adopted the more conservative assessment.

The remainder of this paper is structured as follows. Section~\ref{sec:landscape} surveys the existing policy landscape, identifying what current instruments assume about technical feasibility. Section~\ref{sec:taxonomy} presents the taxonomy of hardware-level governance mechanisms with feasibility assessments. Section~\ref{sec:adversarial} analyses adversarial considerations, circumvention strategies, and structural constraints. Section~\ref{sec:mapping} maps mechanisms to governance scenarios and draws lessons from existing verification regimes. Section~\ref{sec:discussion} discusses implications for the near-term AI governance research agenda.


\section{The Compute Governance Landscape}
\label{sec:landscape}

AI governance proposals increasingly reference compute as a regulatory lever: a proxy for AI capability that is detectable, excludable, and quantifiable in ways that model capabilities and training data are not~\cite{sastry2024computing, heim2024govern}. This section surveys the principal policy instruments that invoke hardware or compute controls, identifying what each \textit{assumes} about technical feasibility without demonstrating it. These unexamined assumptions define the requirements that the taxonomy in Section~\ref{sec:taxonomy} must address.

\subsection{Compute Thresholds in Existing Regulation}

The EU AI Act (Regulation 2024/1689) and US Executive Order~14110 both operationalise compute-based governance through numerical thresholds, but each embeds technical assumptions that remain unverified.

The EU AI Act designates general-purpose AI models trained above $10^{25}$~FLOPs as presumptively possessing ``high impact capabilities,'' classifying them as posing systemic risk and triggering additional obligations beyond those applicable to all general-purpose AI models, including systemic risk assessment, mitigation measures, and serious incident reporting~\cite{euaiact2024, erben2025training}. This threshold mechanism assumes, at minimum, that training compute can be reliably measured or reported, that the threshold meaningfully separates high-risk from lower-risk systems, and that it will remain a useful boundary as the field advances. All three assumptions are problematic. The threshold is static while training compute for frontier systems has doubled approximately every six months, growing by a factor of 350~million over 13~years~\cite{sastry2024computing}. Erben et al.\ recommend adjusting the threshold upward to $10^{26}$~FLOPs and establishing a periodic review mechanism~\cite{erben2025training}, but the Act provides for threshold updates under Article~51(3) without specifying an automatic adjustment formula.

US Executive Order~14110 (October 2023) imposed reporting requirements on developers training above a specified compute threshold and on infrastructure-as-a-service providers hosting large clusters exceeding $10^{20}$~operations per second \textit{and} 100~Gbit/s networking bandwidth~\cite{whitehouse2023eo, heim2024governing}. Its implicit technical assumption was that cloud providers can serve as reliable regulatory intermediaries, capable of identifying and reporting relevant training activity. The Executive Order was revoked by Executive Order~14179 (January 2025)~\cite{byrne2025engineering}, but regardless of its regulatory fate, EO~14110 established an important precedent: it demonstrated that compute-based reporting requirements are administratively implementable and that cloud provider reporting is operationally feasible, even if the accuracy and completeness of such reporting remain untested under adversarial conditions.

The UK has adopted a voluntary approach through the AI Safety Institute (now the AI Security Institute), which conducts pre-deployment evaluations by agreement with developers rather than by statutory mandate~\cite{ritchie2025turing}. This approach avoids the threshold obsolescence problem but assumes continued voluntary participation and does not address the question of how compliance could be verified if participation became contested.

The common thread across all three jurisdictions is a reliance on self-reported compute figures. None has established an independent technical mechanism for verifying that reported training compute is accurate, that relevant training runs have been disclosed, or that compute thresholds have not been circumvented through techniques such as distillation, fine-tuning, or distributed training across reporting boundaries.

\subsection{Export Controls and Supply Chain Instruments}

The US Bureau of Industry and Security's October 2022 and October 2023 restrictions on advanced AI chip exports to China represent what Belfield describes as ``the beginnings of a de facto non-proliferation regime for state-of-the-art AI chips,'' having profoundly changed the governance landscape~\cite{belfield2025domestic, shrivastava2025china}. The controls target chips classified under Export Control Classification Numbers 3A090 and 4A090, using performance thresholds to define restricted items~\cite{kulp2024hardware}, and extend to chipmaking equipment, notably ASML's EUV lithography systems~\cite{cheung2025geopolitical}. Proposed rules have also sought to restrict cloud computing services that would provide equivalent computational access to restricted parties~\cite{kulp2024hardware}.

These controls embed a distinct set of technical assumptions: the ability to verify end-use and end-user compliance after point of sale, the ability to detect diversion through third countries, and the ability to prevent restricted parties from accessing equivalent compute through cloud services or by aggregating lower-performance chips. None of these capabilities is currently provided by the controls themselves. The gap between what export controls assume and what can currently be verified is precisely where hardware-level governance mechanisms become necessary.

\subsection{International Proposals}

At the international level, compute governance proposals remain at the conceptual or early negotiation stage, and their technical assumptions are correspondingly more ambitious. Belfield proposes four interlocking institutions: compute-indexed domestic regulation, an International AI Agency modelled on the IAEA, a Secure Chips Agreement modelled on the NPT, and a US-led allied public-private partnership for frontier AI~\cite{belfield2025domestic}. Ramiah et al.\ propose a ``Global Compute Pause Button'' framework with technical enforcement via FLOP caps, offline licensing, and chip registries~\cite{ramiah2025toward}. Scholefield et al.\ recommend a conditional AI safety treaty with provisions for mutual verification~\cite{scholefield2025international}. Ho et al.\ propose four institutional models, each with precedents in existing international organisations~\cite{ho2023international}.

What unites these proposals is a dependence on technical mechanisms that do not yet exist at governance grade: on-chip compute metering, cryptographic proof of training compliance, hardware-enabled enforcement, and privacy-preserving verification. The policy literature articulates what governance should accomplish; Section~\ref{sec:taxonomy} assesses what technology can currently deliver, and where the critical gaps lie.


\section{Taxonomy of Hardware-Level Governance Mechanisms}
\label{sec:taxonomy}

\begin{figure}[htbp]
	\centering
	\includegraphics[width=\textwidth]{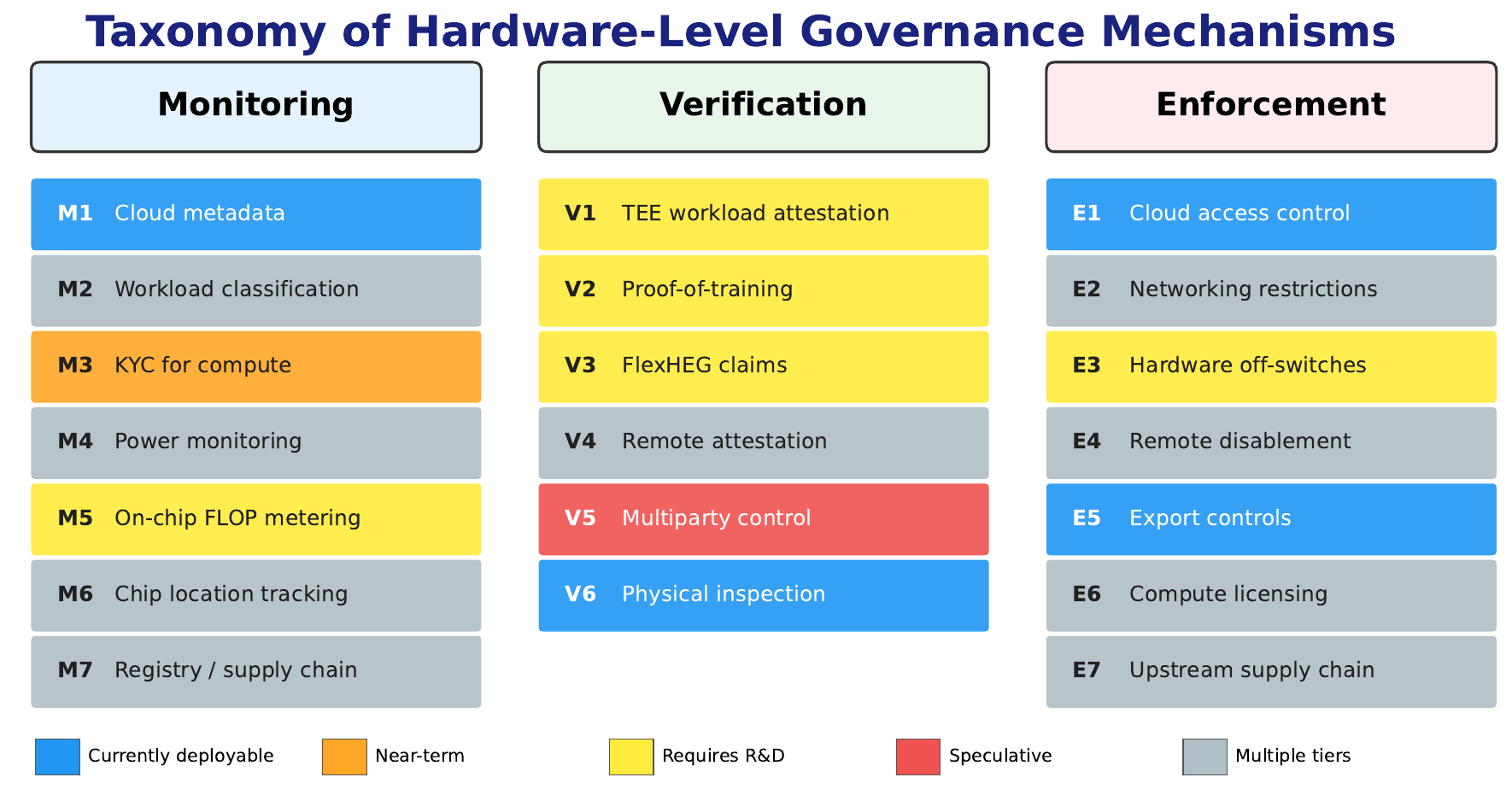}
	\caption{Overview of the 20 hardware-level governance mechanisms organised by function (monitoring, verification, enforcement). Colour indicates the primary feasibility tier from Table~\ref{tab:feasibility}; grey denotes mechanisms that span multiple tiers depending on the implementation variant. Mechanism identifiers (M/V/E) are used throughout Sections~\ref{sec:taxonomy}--\ref{sec:mapping}.}
	\label{fig:taxonomy-overview}
\end{figure}

This section proposes a taxonomy of hardware-level mechanisms that can support AI governance, organised by their primary governance function: monitoring (detecting what compute is being used for), verification (proving compliance claims are true), and enforcement (constraining or penalising non-compliant behaviour). Each mechanism is assigned an identifier (M for monitoring, V for verification, E for enforcement) for cross-referencing with the adversarial analysis in Section~\ref{sec:adversarial} and the scenario mapping in Section~\ref{sec:mapping}. Feasibility is assessed using the four-point scale described in Section~\ref{sec:introduction}. Where a mechanism spans multiple feasibility tiers depending on the implementation variant, both ratings are given. Figure~\ref{fig:taxonomy-overview} provides a visual overview of the full taxonomy.

\subsection{Monitoring Mechanisms}
\label{sec:monitoring}

Monitoring mechanisms provide visibility into how compute resources are being used. Without this visibility, governance instruments such as compute thresholds or training run reporting requirements cannot be verified or enforced. The mechanisms identified range from those already operational within existing cloud infrastructure to speculative hardware modifications that would require years of development.

\paragraph{M1: Cloud provider metadata and billing records.}
The most immediately available monitoring layer exploits information that compute providers already collect for commercial purposes: hardware configuration, chip-hours consumed, types of accelerators allocated, GPU networking topology, power draw, and network bandwidth utilisation~\cite{heim2024governing}; Sastry et al.\ confirm that compute providers' existing data infrastructure covers the core indicators needed for governance monitoring~\cite{sastry2024computing}. Heim et al.\ rate this data layer as ``highly feasible,'' and it is already partially operational through existing provider systems used for billing, service optimisation, and fraud detection~\cite{heim2024governing}. The principal limitation is completeness: metadata collection is confined to individual providers and does not extend to on-premises compute or cross-provider aggregation. \textit{Feasibility: currently deployable.}

\paragraph{M2: Workload classification.}
Building on the metadata in M1, machine learning classifiers can be applied to distinguish AI training workloads from other compute activity. Classifiers trained on datasets such as the MIT Supercloud Dataset have achieved 95\% accuracy in identifying AI training across ten different model architectures, using cluster-level indicators (power consumption, network bandwidth) and node-level indicators (accelerator core utilisation, memory bandwidth)~\cite{heim2024governing}. The RAND game-theoretic analysis additionally identifies the ratio of multiply-accumulate (MAC) operations to total instructions as a distinguishing signal for AI workloads~\cite{moon2025strategies}. While this accuracy is achievable in cooperative cloud environments, adversarial robustness remains undemonstrated: classification accuracy under deliberate evasion, where workloads are structured to mimic non-AI signatures, has not been systematically tested. Classification accuracy also degrades when training is distributed across providers~\cite{krys2025distributed}. \textit{Feasibility: currently deployable (in cooperative environments); near-term (for adversarially robust deployment).}

\paragraph{M3: Know-Your-Customer (KYC) requirements for compute access.}
Analogous to financial KYC obligations, this mechanism requires cloud providers to verify the identity of customers accessing large-scale compute, including beneficial ownership, key personnel, and stated purpose~\cite{egan2023oversight}. US Executive Order~14110 imposed identity verification requirements on US-based infrastructure-as-a-service providers for foreign customers~\cite{whitehouse2023eo}, establishing a functional precedent for KYC-style obligations in the compute sector. The mechanism is technically straightforward at the individual provider level; the challenge lies in cross-provider coordination. A customer denied access by one provider can fragment their compute across multiple providers or jurisdictions. Egan and Heim propose threshold-based monitoring where providers track chip-hour accumulation and set warnings at defined fractions of regulatory thresholds, without accessing customer data or underlying code~\cite{egan2023oversight}. The RAND game-theoretic analysis notes that KYC is not dependent on compute thresholds but rather on access to specialised metrics that reveal user identity and intent~\cite{moon2025strategies}. \textit{Feasibility: near-term.}

\paragraph{M4: Power consumption monitoring.}
Large-scale AI training runs produce distinctive sustained high-power profiles. The BLOOM model, for instance, consumed approximately 3,600~kWh per day during training~\cite{moon2025strategies}. At the facility level, power consumption is already metered and can serve as a coarse proxy for compute activity. For international agreements, energy monitoring can detect large GPU clusters and convert consumption estimates into approximate FLOP figures~\cite{wasil2024verification}. Barnett et al.\ discuss electricity monitoring as a verification mechanism for fab shutdown~\cite{barnett2025technical}. Remote sensing via infrared imaging can also detect undeclared data centres through heat signatures~\cite{wasil2024verification}. The limitation is specificity: power monitoring alone cannot distinguish AI training from other high-performance computing workloads such as climate simulation, molecular dynamics, or financial modelling without supplementary signals. No published study quantifies false positive rates for power-based training detection; establishing these error bounds is a necessary step before power monitoring can underpin regulatory decisions. \textit{Feasibility: currently deployable (for coarse detection); near-term (for AI-specific identification when combined with other signals).}

\paragraph{M5: On-chip compute metering.}
Hardware-level mechanisms that count floating-point operations (FLOPs) performed by an AI accelerator represent the most technically ambitious monitoring proposal. Several architectures have been proposed. Petrie describes embedded security blocks, each requiring fewer than 40,000~transistors, that track a ``usage allowance'' which decrements with each authorised computation~\cite{petrie2025embedded}. At 10,000~blocks per chip, the total die area overhead on an NVIDIA H100 (80~billion transistors) would be less than~1\%. Petrie et al.\ extend this to FlexHEG (Flexible Hardware-Enabled Guarantee) designs, where an auxiliary ``guarantee processor'' within a tamper-proof enclosure monitors all instructions and data transfers to and from the AI chip, enabling FLOP tracking that prevents double-counting~\cite{petrie2025flexible}. Ramiah et al.\ propose tamper-proof FLOP caps that automatically halt execution once a predefined threshold is reached~\cite{ramiah2025toward}. The RAND analysis of hardware-enabled governance mechanisms notes that NVIDIA H100 GPUs already contain on-device performance counters tracking FLOPs, instruction counts, NVLink bandwidth, PCIe bandwidth, memory accesses, and power consumption; these existing counters could be adapted for governance purposes, with additional logic to set maximum values determined by a licence~\cite{kulp2024hardware}. However, Barnett et al.\ rate on-chip mechanisms at ``low technological readiness,'' noting the absence of functional prototypes~\cite{barnett2025technical}. Aarne et al.\ estimate that full implementation would require 18~months to 4~years of development by leading chip firms, with a further 4~years for sufficiently widespread deployment~\cite{aarne2024secure}. \textit{Feasibility: requires R\&D.}

The emerging literature presents three distinct architectures for on-chip metering. \textit{Architecture~A} (Petrie's embedded off-switches) distributes security across thousands of small blocks, each independently gating essential logic and verifying licences using compact ECDSA circuitry, with public keys stored in MaskROM and nonces generated from ring-oscillator TRNGs~\cite{petrie2025embedded}. Its strength is redundancy; its weakness is that each block is physically exposed across the chip surface, with no single tamper-proof boundary. \textit{Architecture~B} (labelled FlexHEG in Petrie et al.) concentrates enforcement in a single auxiliary guarantee processor inside a tamper-proof enclosure; Petrie et al.\ recommend open-source root-of-trust designs such as OpenTitan or Caliptra for this processor~\cite{petrie2025flexible}. This is easier to harden physically but creates a single point of failure; since the guarantee processor sits between the AI chip and the outside world, the performance cost of mediating all chip I/O is an open engineering question not yet addressed in the published literature. \textit{Architecture~C} adapts existing on-device performance counters, the lowest-cost path since it builds on existing silicon~\cite{kulp2024hardware}; in our assessment this approach is the weakest against adversarial actors, as current performance counters are accessible via firmware interfaces and were designed for debugging rather than tamper-resistant governance~\cite{kulp2024hardware}. The comparative picture suggests a hybrid strategy: repurposed counters (C) for near-term domestic regulation where providers are cooperative, and enclosed guarantee-processor designs (B) for higher-assurance international verification.

\paragraph{M6: Chip location tracking.}
Verifying the physical location of AI chips is critical for export control enforcement and treaty compliance~\cite{belfield2025domestic}. Delay-based geolocation measures the round-trip time of cryptographic challenge-response exchanges between a chip and trusted landmark servers, exploiting the physical constraint that signals cannot travel faster than the speed of light~\cite{brass2024location}. Brass and Aarne provide the most detailed analysis, concluding that pure-software delay-based solutions are ``both feasible and relatively cheap to implement on chips in the near future,'' estimating setup and maintenance costs below \$1,000,000 over several years, with a network of ``several dozen'' landmark servers~\cite{brass2024location}. NVIDIA H100 GPUs already contain trusted execution environments that could implement location attestation~\cite{brass2024location}. Petrie et al.\ describe FlexHEG-based location verification with privacy-preserving features~\cite{petrie2025flexible}. Spoofing resistance is stronger than GPS or RFID, since an adversary would need to relay signals at light speed, but dedicated spoofing infrastructure at scale remains a theoretical concern. \textit{Feasibility: near-term (for software-based delay methods); requires R\&D (for hardware-integrated solutions with strong tamper resistance).}

\paragraph{M7: Chip registry and supply chain tracking.}
Assigning unique identifiers to state-of-the-art AI chips and maintaining a registry tracking their lifecycle from fabrication to deployment to destruction provides a chain-of-custody layer analogous to nuclear material accountancy~\cite{belfield2025domestic, baker2023nuclear}. Arslan discusses physical unclonable functions (PUFs) as a potential technique for secure chip identification, though noting that further research on security and feasibility is needed~\cite{arslan2025advancing}. Ramiah et al.\ propose blockchain-based chip serialisation for a global compute supply registry~\cite{ramiah2025toward}. At the trade data level, Ramli demonstrates that existing customs data (HS codes 8542 and 8419) can track semiconductor flows using physicality correlation frameworks, counterfactual modelling, and Granger causality tests to identify shadow supply chains and intermediary jurisdictions used to circumvent export controls~\cite{ramli2026pattern}. The concentration of advanced chip manufacturing (fewer than two dozen facilities for sub-14nm chips as of 2023) makes registry enforcement more tractable than it would be for a commodity with distributed production~\cite{shavit2023does}. \textit{Feasibility: near-term (for registry systems); currently deployable (for trade data analysis).}

\subsection{Verification Mechanisms}
\label{sec:verification}

Verification mechanisms enable a party (a regulator, treaty body, or auditor) to confirm that governance-relevant claims about compute usage are true without necessarily requiring direct access to proprietary model details. This is the most technically demanding category, as it must satisfy two competing requirements: providing sufficient evidence of compliance while preserving the confidentiality of commercial and national security interests.

\paragraph{V1: Trusted execution environments (TEEs) for workload attestation.}
TEEs create isolated processing environments where code and data are protected from the host operating system, hypervisor, and even the hardware owner~\cite{ccc2022technical}; they are built into modern CPUs and increasingly into AI accelerators, notably the NVIDIA H100~\cite{kulp2024hardware}. In a governance context, a developer (the ``attester'') could provably demonstrate properties of their workload to a regulator (the ``verifier'') without revealing model architecture, training data, or weights. Verifiable properties might include: that a specific safety evaluation was run and produced a specific result, that a prohibited dataset was not used during training, or that total training compute did not exceed a threshold~\cite{heim2024governing, petrie2025flexible}. Heim et al.\ rate detailed workload verification via TEEs as ``currently not possible without directly observing customer code or data,'' but note that confidential computing techniques are approaching this capability~\cite{heim2024governing}. Petrie et al.\ note that current GPU confidential computing implementations do not yet support multi-GPU workloads, but that NVIDIA has suggested adding such support in the next generation~\cite{petrie2025flexible}. The Confidential Computing Consortium identifies key limitations: hardware-based cryptographic primitives lack crypto-agility (if a primitive is deprecated, hardware replacement is required), and TEEs are vulnerable to side-channel attacks through power analysis and timing analysis~\cite{ccc2022technical}. \textit{Feasibility: requires R\&D (for governance-grade deployment on AI accelerators).}

Existing TEEs differ substantially in both isolation model and governance relevance. Intel SGX provides process-level enclave isolation, whereas Intel TDX shifts toward VM-level isolation for entire trust domains; for governance purposes, TDX is more relevant because it can in principle isolate a whole training VM. But SGX's history is cautionary: repeated side-channel and fault-injection vulnerabilities (Spectre-class issues, Foreshadow, Plundervolt, SGAxe) illustrate how difficult it is to make enclave security robust in practice. AMD SEV-SNP offers VM-level isolation but is less directly relevant to frontier AI governance because its attestation surface sits primarily on the CPU side. ARM's Confidential Compute Architecture is best understood as an edge case for governance: potentially useful for inference-time monitoring on edge devices, but not a near-term basis for datacentre-scale frontier training. Across these vendor families, the Confidential Computing Consortium's taxonomy usefully distinguishes hardware TEEs, trusted platform modules, attestation flows, and hardware roots of trust~\cite{ccc2022technical}.

For AI governance specifically, NVIDIA's Confidential Computing implementation on Hopper H100 is currently the most relevant because the TEE is located on the AI accelerator itself. The RAND HEM report describes a root-of-trust anchored in keys fused onto the GPU at manufacture, secure boot that verifies signed firmware, and attestation reports verified by a CPU TEE~\cite{kulp2024hardware}. Brass and Aarne likewise identify the H100 TEE as a plausible substrate for location attestation~\cite{brass2024location}. The central limitation is scale: current GPU confidential computing implementations do not yet support multi-GPU workloads~\cite{petrie2025flexible}, whereas frontier training runs span hundreds or thousands of accelerators. NVIDIA has indicated that broader multi-GPU support will be added in a future product generation~\cite{petrie2025flexible}, but until cluster-wide confidential computing is production-ready, TEEs are more suited to bounded, single-GPU verification tasks than to attesting full frontier training runs.

Historical experience with TEEs in embedded and adversarial deployments cautions against assuming that security properties demonstrated in datacentre-oriented threat models will carry over to edge settings. Recent work on TrustZone-assisted TEEs shows continued architectural flaws and kernel-level crashes under systematic testing~\cite{wen2025teedfuzzer}. Symbolic validation of low-end microcontroller enclaves shows that automated tools can autonomously rediscover interface-sanitization vulnerabilities in embedded TEEs that were previously identified and patched only through manual analysis~\cite{goossens2025principled}.

\paragraph{V2: Cryptographic proof-of-learning and proof-of-training.}
Jia et al.\ introduced Proof-of-Learning (PoL), a mechanism that exploits the stochasticity of stochastic gradient descent to create verifiable certificates that a specific model was trained through a specific computational process~\cite{jia2021proof}. The prover periodically logs weight snapshots and data indices during training; a verifier can then replay segments to confirm consistency. Shavit extends this to a regulatory context, proposing on-chip firmware that periodically saves weight snapshots, computes cryptographic hashes, and stores or transmits them to a verifier-trusted server~\cite{shavit2023does}. Inspectors then analyse logs from sampled chips using ``Proof-of-Training-Transcript'' (PoTT) protocols. Shavit notes that zero-knowledge proofs are computationally inefficient at this scale, and proposes instead that the prover and verifier agree on a neutral, jointly-trusted cluster on which verification protocols can be executed~\cite{shavit2023does}. The RAND game-theoretic model recommends pairing compute-based detection methods with non-compute-based governance approaches such as KYC~\cite{moon2025strategies}. The principal limitation is that these protocols remain nascent: Shavit acknowledges that developing provably secure and efficient PoTT protocols is ``an important avenue for future work''~\cite{shavit2023does}. \textit{Feasibility: requires R\&D.}

\paragraph{V3: FlexHEG-based verifiable claims.}
The FlexHEG architecture~\cite{petrie2025flexible} proposes an auxiliary guarantee processor within a tamper-proof enclosure that can produce verifiable records of training activity. The guarantee processor encrypts and authenticates data from the AI chip, and tamper-detection systems trigger secret-wiping mechanisms (including permanent fuse-blowing) if physical intrusion is attempted. The architecture supports verification of: compute quantity used (FLOP tracking with double-counting prevention), training data used (via hashes), model properties and training techniques, deployment conditions, and negative claims (proving that certain operations were \textit{not} performed) via compute accounting. The authors recommend open-source IP blocks such as OpenTitan or Caliptra for the guarantee processor to enable public auditability of the verification logic. They state: ``We are confident that some variant of this proposal is technically feasible. However, early versions may need to compromise on security and the sophistication of supported rules''~\cite{petrie2025flexible}. Achieving high physical security standards ``could take several years''~\cite{petrie2025flexible}. \textit{Feasibility: requires R\&D.}

\paragraph{V4: Remote attestation via cryptographic licensing.}
A specialised co-processor on an AI chip holds a cryptographically signed digital certificate; authorisation is periodically renewed by a regulatory authority, and expiry or invalidity causes the chip to degrade performance or halt operation~\cite{petrie2025flexible, belfield2025domestic, aarne2024secure}. Petrie provides a detailed circuit-level design: ECDSA verification implemented in approximately 9,000~gates ($\sim$40,000~transistors per security block), nonces generated by on-chip true random number generators using ring oscillators ($\sim$70~transistors), and public keys hardwired into MaskROM~\cite{petrie2025embedded}. The RAND HEM report describes NVIDIA's existing Confidential Computing functionality on the H100 as including a hardware root of trust anchored in keys fused onto the GPU at manufacture, secure boot that verifies signed firmware, and remote attestation capabilities verified by a CPU TEE~\cite{kulp2024hardware}. Sastry et al.\ describe hardware-based remote enforcement as ``highly speculative'' and ``unproven''~\cite{sastry2024computing}. The gap between existing attestation infrastructure (commercially deployed for multi-tenant cloud security) and governance-grade attestation (resistant to nation-state adversaries with physical access) remains substantial but is narrowing. \textit{Feasibility: requires R\&D (for governance-grade adversarial resistance); near-term (for building on existing commercial TEE infrastructure).}

\paragraph{V5: Multiparty cryptographic control of training runs.}
Proposals for distributing control over training run initiation converge on a similar goal but use somewhat different framings. Sastry et al.\ describe a mechanism using multisignature cryptographic protocols where training only executes when cryptographically signed by all required parties~\cite{sastry2024computing}; Belfield describes multiparty control where large training runs must be cryptographically approved by all parties before they begin~\cite{belfield2025domestic}; Petrie et al.\ propose that operating licences require approval from all or a majority of a defined set of governments~\cite{petrie2025flexible}. Conceptually, this implements a ``two-person rule'' analogous to nuclear launch protocols. No prototype exists; major open questions include key management architecture, timeout handling when a signatory is unresponsive, governance structure for signatory selection, and the latency implications of adding cryptographic approval to training initiation pipelines. \textit{Feasibility: speculative.}

\paragraph{V6: Physical inspection and on-site auditing.}
Periodic or surprise inspections of data centres, fabrication facilities, and chip storage sites remain the most straightforward verification mechanism and the one with the strongest precedent from existing verification regimes. Barnett et al.\ rate periodic data centre inspections at high technological readiness, identifying no technological hurdles to implementation~\cite{barnett2025technical}. Brundage et al.\ describe a four-level assurance framework for AI auditing, where the highest level (AAL-4) includes short-notice inspections of physical facilities, hardware attestation and verification, destructive testing of hardware samples, and continuous monitoring designed to detect active deception; however, the authors note that these highest assurance levels are ``not yet technically and organizationally feasible''~\cite{brundage2026frontier}. The IAEA model is a frequently cited reference point: it combines material accountancy (tracking declared nuclear materials through facilities), on-site inspections, remote surveillance, environmental sampling, and satellite imagery~\cite{baker2023nuclear, ramiah2025toward}. Wasil et al.\ categorise physical inspection methods as requiring some additional research but no fundamental technological breakthroughs~\cite{wasil2024verification}. The limitation is scalability: physical inspections are labour-intensive and cannot provide continuous monitoring. \textit{Feasibility: currently deployable.}

\subsection{Enforcement Mechanisms}
\label{sec:enforcement}

Enforcement mechanisms constrain non-compliant behaviour, either proactively (preventing violations before they occur) or reactively (imposing consequences after detection). The spectrum ranges from administrative controls already in use to hardware-embedded enforcement that remains largely theoretical.

\paragraph{E1: Cloud provider access control.}
Compute providers can deny or limit access to compute resources for non-compliant customers by suspending accounts, restricting accelerator allocation, or refusing to deploy unlicensed models~\cite{heim2024governing}. This is ``highly feasible'' and ``widely used'' in existing cloud operations~\cite{heim2024governing}. US Executive Order~14110 required infrastructure-as-a-service providers to verify the identity of foreign customers and to submit reports when foreign persons used services to train large AI models~\cite{whitehouse2023eo}. The principal limitation is jurisdictional: customers can migrate to providers in less regulated jurisdictions, and on-premises compute is entirely outside this mechanism's reach. \textit{Feasibility: currently deployable.}

\paragraph{E2: Chip-to-chip networking restrictions.}
Physically or logically limiting chip-to-chip networking bandwidth to prevent chips from being aggregated into clusters large enough for frontier training is one of the more frequently discussed but technically contentious proposals. Aarne et al.\ discuss a proposal, originating from a US Bureau of Industry and Security request for comment, to restrict inter-chip communication to less than 1~GB/s outside a pre-authorised pod of chips~\cite{aarne2024secure}. Kulp et al.\ propose a ``Fixed Set'' mechanism where high-bandwidth communication would be limited to a selected number of preauthorised chips in a pod, with restricted bandwidth outside; while the H100's NVLink already enables high-bandwidth intra-pod communication, the governance-enforced bandwidth restriction itself does not yet exist~\cite{kulp2024hardware}. Chips designed with inherent networking limits ``do not exist today, but could potentially be developed''~\cite{heim2024govern}. Critically, Ramiah et al.\ warn that the ``long-term utility is increasingly questionable due to rapid progress in distributed training methods''~\cite{ramiah2025toward}. Low-communication training algorithms are advancing rapidly, with startups already pre-training models up to 10~billion parameters using decentralised methods across continents at ordinary internet speeds~\cite{krys2025distributed}. Quantitative analysis of what bandwidth thresholds would prevent frontier training while permitting legitimate high-performance computing remains absent from the literature. \textit{Feasibility: speculative (for chip-level restrictions); near-term (for external infrastructure limits on cluster size).}

\paragraph{E3: Hardware-embedded off-switches.}
Petrie provides the most detailed design for hardware kill switches: thousands of independent security blocks distributed across a chip's die, each gating a piece of essential logic and acting as a deadman's switch~\cite{petrie2025embedded}. Each block requires periodic cryptographic authorisation (a ``licence'') to continue operating; if the usage allowance reaches zero, the block halts its controlled logic, introducing faults that render the chip non-functional. Petrie et al.\ extend this to FlexHEG enforcement, where tamper-detection systems can permanently disable a chip by blowing microscopic fuses if physical intrusion is detected~\cite{petrie2025flexible}. Petrie et al.\ propose approval structures requiring all or a majority of a defined set of governments before certain operations proceed, which could reduce single-party control of the kill capability~\cite{petrie2025flexible}. The RAND report describes an ``Offline Licensing'' scheme where GPUs fall back to 1\% of computational performance upon licence expiration, noting similarity to Intel's commercially viable ``Intel On Demand'' feature-gating product~\cite{kulp2024hardware}. Hardware development timelines remain a central unresolved constraint: ``Hardware typically takes years to design and manufacture, and longer still for new AI accelerators to displace older ones''~\cite{petrie2025embedded}. \textit{Feasibility: requires R\&D.}

\paragraph{E4: Remote performance degradation and disablement.}
A variant of the off-switch concept, remote disablement enables ongoing enforcement rather than one-time point-of-sale controls. If a chip's cryptographic licence is not renewed, performance degrades or operation ceases, effectively ``digitising'' export controls~\cite{sastry2024computing}. If chips are diverted to an unauthorised jurisdiction, the licence signal can be withheld~\cite{belfield2025domestic}. Brass and Aarne describe ``region locking'' that disables chip functionality if location cannot be verified~\cite{brass2024location}. The principal objection is political, not technical: concentrating remote-disable authority creates a single point of geopolitical leverage that many states will find unacceptable. Governance architectures that distribute this authority (e.g., requiring consensus among multiple independent jurisdictions to trigger disablement) have been proposed conceptually but not designed in detail. \textit{Feasibility: requires R\&D (technically); speculative (politically, due to sovereignty concerns).}

\paragraph{E5: Export controls on hardware.}
The existing regime of restricting sale and transfer of advanced AI chips to designated jurisdictions, implemented through the US Bureau of Industry and Security's Entity List and Export Control Classification Numbers (ECCN 3A090, 4A090), is currently the most operationally implemented hardware-level enforcement mechanism~\cite{belfield2025domestic, sastry2024computing}. Controls extend to chipmaking equipment, notably EUV lithography systems manufactured by ASML. Belfield assesses the regime as ``mostly successful'' in curbing proliferation of cutting-edge chips, though circumvention via third-country re-export is documented~\cite{belfield2025domestic}. Ramli demonstrates that trade data analysis can detect evasion through shadow supply chains~\cite{ramli2026pattern}. The mechanism's efficacy is structurally dependent on manufacturing concentration: as of 2023, advanced logic chip fabrication is dominated by TSMC (Taiwan), Samsung (South Korea), and Intel (US). Should this concentration erode, the leverage of export controls would diminish correspondingly. \textit{Feasibility: currently deployable.}

\paragraph{E6: Compute licensing and registration.}
Requiring regulatory licences before conducting training runs above specified compute thresholds is the enforcement mechanism most directly embedded in existing law. The EU AI Act establishes a $10^{25}$~FLOP threshold for classifying general-purpose AI models as posing systemic risk, triggering additional obligations on top of the baseline requirements that apply to all general-purpose AI models~\cite{euaiact2024}. US Executive Order~14110 established reporting requirements above a threshold~\cite{whitehouse2023eo}. Pistillo and Villalobos identify four loopholes that allow circumvention: fine-tuning below the threshold to enhance a pre-trained model, model reuse techniques (knowledge distillation, kickstarting, and reincarnation) that transfer capabilities from larger models, model expansion that grows a model's parameters beyond the threshold after initial training below it, and inference-time compute scaling that shifts capability-enhancing computation from training to deployment~\cite{pistillo2025defending}. Petrie et al.\ propose hardware-enforced licensing via FlexHEG operating licences: time-limited cryptographic licences that function as remote off-switches if not renewed~\cite{petrie2025flexible}. \textit{Feasibility: currently deployable (for regulatory reporting); requires R\&D (for hardware-enforced licensing).}

\paragraph{E7: Upstream supply chain controls.}
Controlling access to chipmaking equipment (EUV lithography machines, photomasks, high-purity silicon wafers) and monitoring fabrication facility construction and operation represents the most upstream enforcement point. Barnett et al.\ rate monitoring of new fab construction at high technological readiness, and separately note that satellite imagery and electricity consumption data can verify whether existing facilities have been shut down~\cite{barnett2025technical}. ASML export restrictions on EUV equipment are already in force. Ramiah et al.\ describe production controls modelled on Good Manufacturing Practice (GMP) frameworks from pharmaceutical regulation~\cite{ramiah2025toward}. Wasil et al.\ note that fab inspections, like data centre inspections, require some additional research but no fundamental new technology~\cite{wasil2024verification}. The analogy to nuclear governance is suggestive: just as non-proliferation regimes restrict access to enrichment and reprocessing technology, AI governance could restrict access to advanced lithography, though the parallel is imperfect and neither Baker nor Vermeer draws this specific comparison~\cite{vermeer2024historical}. \textit{Feasibility: currently deployable (for monitoring and equipment restrictions); near-term (for comprehensive production control frameworks).}

\subsection{Feasibility Summary}
\label{sec:feasibility-summary}

Table~\ref{tab:feasibility} summarises the 20 mechanisms by feasibility tier. Several mechanisms span multiple tiers depending on the implementation variant (e.g., M6 chip location tracking is near-term for software-based delay methods but requires R\&D for hardware-integrated solutions; E6 compute licensing is currently deployable for regulatory reporting but requires R\&D for hardware enforcement). The distribution reveals a structural feature of the current governance landscape: most mechanisms with high feasibility provide monitoring or administrative enforcement capabilities, while the mechanisms most needed for international treaty verification (M5 on-chip metering, V2 proof-of-training, E3 hardware off-switches) remain in the R\&D stage. This gap between what governance frameworks demand and what technology can currently deliver defines the near-term research agenda for hardware-level AI governance.

\begin{table}[htbp]
\centering
\caption{Feasibility classification of hardware-level governance mechanisms. Mechanism identifiers (M/V/E) correspond to the descriptions in Sections~\ref{sec:monitoring}--\ref{sec:enforcement}. Some mechanisms appear at multiple tiers reflecting variant-dependent readiness.}
\label{tab:feasibility}
\small
\begin{tabular}{p{2.8cm} p{3.8cm} p{3.8cm} p{3.8cm}}
\toprule
\textbf{Feasibility Tier} & \textbf{Monitoring} & \textbf{Verification} & \textbf{Enforcement} \\
\midrule
Currently deployable & M1 Cloud metadata; M2 Workload classification; M4 Power monitoring; M7 Trade data analysis & V6 Physical inspection & E1 Cloud access control; E5 Export controls; E6 Compute licensing (reporting) \\
\addlinespace
Near-term & M2 Workload classification; M3 KYC; M6 Location tracking (software); M7 Chip registry & V4 Remote attestation (commercial TEEs) & E2 Networking limits (external); E7 Upstream supply chain controls \\
\addlinespace
Requires R\&D & M5 On-chip FLOP metering & V1 TEE workload attestation; V2 Proof-of-training; V3 FlexHEG claims; V4 Remote attestation (governance-grade) & E3 Hardware off-switches; E4 Remote disablement; E6 Compute licensing (hardware-enforced) \\
\addlinespace
Speculative & & V5 Multiparty cryptographic control & E2 Networking limits (on-chip); E4 Remote disablement (political feasibility) \\
\bottomrule
\end{tabular}
\end{table}

\begin{figure}[htbp]
	\centering
	\includegraphics[width=\textwidth]{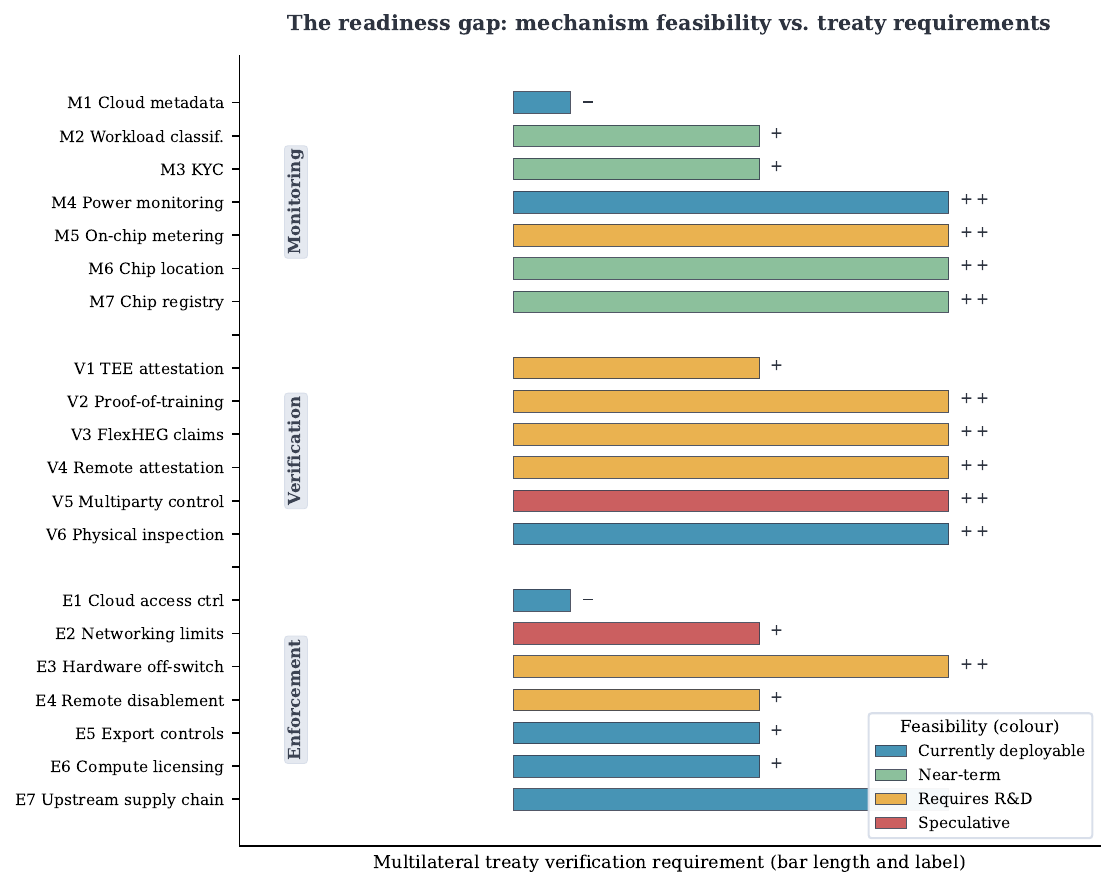}
	\caption{The readiness gap between mechanism feasibility and multilateral treaty requirements. Bar length and label indicate each mechanism's importance for the treaty verification scenario (from Table~\ref{tab:mapping}); bar colour indicates feasibility tier (from Table~\ref{tab:feasibility}). The structural mismatch is visible: mechanisms rated as strong fits ($++$) for treaty verification are predominantly amber (requires R\&D) or red (speculative), while currently deployable mechanisms (blue) have limited treaty applicability.}
	\label{fig:readiness-gap}
\end{figure}

Four mechanisms merit particular attention for near-term R\&D investment: M5 on-chip compute metering, which provides the foundation for all threshold-based governance; V2 proof-of-training protocols, which enable privacy-preserving compliance verification; V3 FlexHEG-based verifiable claims, which offer the most comprehensive verification architecture; and E6 hardware-embedded licensing, which extends enforcement beyond the point of sale. The convergence of these four lines of development would, for the first time, provide the technical substrate for verifiable international agreements on AI development.


\section{Feasibility Constraints and Adversarial Considerations}
\label{sec:adversarial}

A taxonomy of governance mechanisms is only useful if it honestly assesses what can be defeated, circumvented, or rendered obsolete. This section examines the principal classes of threat to the mechanisms described in Section~\ref{sec:taxonomy}, organised from the most immediate and well-documented to the more structural and long-term. An honest treatment of these constraints is essential: policy proposals that ignore adversarial dynamics will either fail on deployment or, worse, create a false sense of security that delays the development of more robust alternatives.

\subsection{Algorithmic Efficiency and the Erosion of Compute Thresholds}
\label{sec:algorithmic-efficiency}

The most widely discussed structural threat to compute-based governance is that algorithmic progress continuously reduces the compute required to achieve any given level of AI capability. Sastry et al.\ note that each year it becomes ``more feasible to train models to a given level of performance using less, cheaper, and more decentralized compute, and consequently somewhat less governable''~\cite{sastry2024computing}. Hooker formalises this concern, arguing that ``the relationship between compute and risk is highly uncertain and rapidly changing'' and that hard-coded thresholds are therefore ``shortsighted and likely to fail to mitigate risk''~\cite{hooker2024limitations}.

The problem manifests through several specific channels. First, inference-time compute scaling (chain-of-thought prompting, best-of-n sampling, retrieval-augmented generation) can dramatically alter a model's risk profile without any additional training compute, a dimension that current thresholds do not capture~\cite{hooker2024limitations}. Pistillo and Villalobos treat inference as a loophole category, noting that above-compute-optimal inference can enhance capabilities beyond what training compute alone would predict~\cite{pistillo2025defending}. Second, knowledge distillation and model reuse allow developers to obtain models with comparable capabilities using substantially less compute, potentially keeping them below regulatory thresholds entirely~\cite{pistillo2025defending}. Third, fine-tuning requires typically less than 1\% of pre-training compute, making it ``very widespread and thus impractical for regulators to track''~\cite{pistillo2025defending}. Pistillo and Villalobos demonstrate that a developer could train a ``teacher model'' above the threshold, never deploy it, and use distillation to produce a marketable model below the threshold~\cite{pistillo2025defending}.

The ``efficiency shock'' described by Alexander compounds these concerns: open-weight models with strong reasoning capabilities can now be produced at a fraction of the compute used by their predecessors, rendering the EU AI Act's $10^{25}$~FLOP threshold ``structurally obsolete'' for defining systemic risk~\cite{alexander2026efficiency}. Once released, open-weight models cannot be recalled or patched, and provenance data can be stripped or forged~\cite{alexander2026efficiency}. Puri further demonstrates that models well below governance thresholds can pose significant safety risks, and that ``naively adjusting compute metering thresholds to block attacks from miniaturised AI systems would significantly disrupt many non-nefarious academic and business AI use cases''~\cite{puri2026small}.

The implication for hardware-level governance is that mechanisms tied exclusively to training compute (M5 FLOP caps, E6 training-time licensing, M2 workload classification) will require continuous threshold adjustment and must eventually be supplemented by capability-based or output-based evaluation. Casper et al.\ propose seven principles for compute accounting standards that reduce gaming, but acknowledge that ``technical ambiguities in how to perform this accounting create loopholes that can undermine regulatory effectiveness''~\cite{casper2025practical}. Jones identifies additional structural problems: threshold updates lag behind algorithmic advances, the update process itself is slow and subject to procedural hurdles, and the definition of a ``distinct model'' becomes unclear for aggregated systems or iteratively refined prototypes~\cite{jones2025overcoming}.

\subsection{Distributed and Decentralised Training}
\label{sec:distributed}

The assumption underlying most monitoring and enforcement mechanisms in Section~\ref{sec:taxonomy} is that frontier AI training occurs in centralised, identifiable data centres using large, interconnected clusters of high-end accelerators. This assumption is increasingly fragile. Krys et al.\ document that low-communication training algorithms now make it ``potentially feasible to conduct training runs over typical Internet speeds,'' with startups having ``already managed to pre-train models up to 10 billion parameters using decentralised methods across continents''~\cite{krys2025distributed}. While it remains ``unclear to what extent low-communication training can keep up with the frontier''~\cite{krys2025distributed}, the trajectory is unfavourable for governance mechanisms that depend on cluster-level visibility.

Distributed training creates three specific governance challenges. First, \textit{compute structuring}: analogous to financial structuring used to evade anti-money-laundering rules, a malicious actor can divide a training run across multiple cloud providers to keep each segment below reporting thresholds~\cite{heim2024governing}. The RAND game-theoretic model demonstrates that BLOOM's $5.8 \times 10^{23}$~FLOPs training could theoretically be hidden as ``58 distinct instances of $10^{22}$~FLOPs''~\cite{moon2025strategies}. Detection without information sharing between providers is ``challenging''~\cite{krys2025distributed}. Second, decentralised peer-to-peer training creates a \textit{no off-switch problem}, as there is no single point of control that a regulator can target~\cite{krys2025distributed}. Third, \textit{communication obfuscation}: strategies to mask inter-node communication patterns are ``likely possible,'' with INTELLECT-1's infrequent synchronisation (once every 38 minutes) cited as an example of how training signatures can be disguised~\cite{krys2025distributed}.

Networking restrictions (E2) have been proposed as a countermeasure to distributed training, but Ramiah et al.\ warn that their utility is ``increasingly questionable,'' citing research demonstrating that ``large-scale training can now be distributed with two orders of magnitude less interconnect bandwidth''~\cite{ramiah2025toward}. Mueller argues more broadly that AI compute is ``just distributed computing'' and that governing it as a general capability would require total control of distributed computing infrastructure, which he describes as at best unrealistic~\cite{mueller2025s}. This is the strongest sceptical position in the literature and one that proponents of compute governance must engage with directly: if distributed training continues to scale, the physical detectability of frontier training runs, the foundational premise of compute governance, could erode within a decade.

\subsection{Physical and Side-Channel Attacks on Hardware Mechanisms}
\label{sec:physical-attacks}

Hardware-embedded governance mechanisms (M5, V3, E3) face a threat model that is qualitatively different from software-based governance: the adversary has physical possession of the device. Petrie catalogues the principal attack vectors against embedded off-switches: logical flaws in design, licence reuse, execution bypass through workload modification, voltage or laser glitching, physical tampering (including focused ion beam editing), secret extraction via probing or side-channel analysis, supply chain compromise, and cryptographic vulnerabilities including the long-term threat from quantum computing~\cite{petrie2025embedded}.

The critical analytical move, often elided in the governance literature, is to tier adversaries explicitly rather than treating ``tamper-proof'' as a binary property. We distinguish three tiers, summarised in Figure~\ref{fig:adversary-tiers}:

\begin{figure}[htbp]
	\centering
	\includegraphics[width=\textwidth]{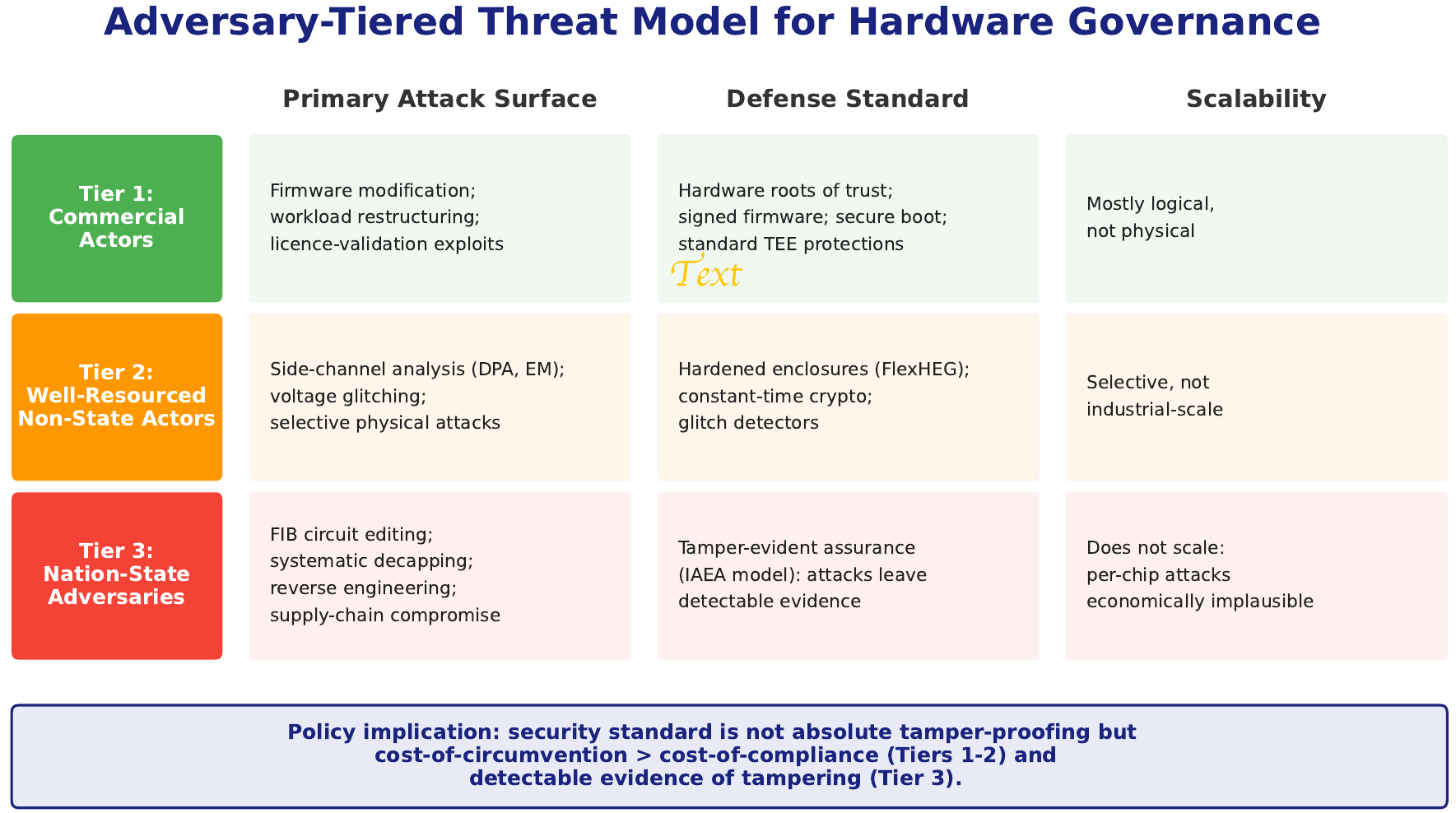}
	\caption{Adversary-tiered threat model for hardware-embedded governance mechanisms. Each tier is characterised by its primary attack surface, the corresponding defense standard, and the scalability of attacks. The policy implication is that tamper resistance need not be absolute: it must make circumvention more expensive than compliance for Tiers~1--2 and leave detectable evidence for Tier~3.}
	\label{fig:adversary-tiers}
\end{figure}

\paragraph{Tier 1: Commercial actors} seeking to evade licensing costs or reporting requirements. Their realistic attack surface is mostly logical rather than physical: modifying firmware where secure boot is weak, restructuring workloads to evade classifiers, or exploiting flaws in licence-validation logic. These attacks are addressable, in principle, with current techniques such as hardware roots of trust, signed firmware, and properly enforced secure boot chains, all of which already exist in some form on high-end accelerators such as the H100~\cite{kulp2024hardware, aarne2024secure}. However, the RAND HEM report notes that these existing features were designed for cloud multi-tenancy and ``do not protect the operation of the GPU itself from physical attack''~\cite{kulp2024hardware}; protections focus on making circumvention incrementally more costly rather than fully preventing it.

\paragraph{Tier 2: Well-resourced non-state actors} operating in permissive jurisdictions. Physical attacks become realistic but still selective rather than industrial. Differential power analysis or electromagnetic side-channel attacks against compact ECDSA or AES implementations are within reach of commercially available tooling, as are voltage-glitch attacks aimed at corrupting licence verification or allowance counters. This matters because the distributed-block proposals (Architecture~A in Section~\ref{sec:monitoring}) rely on extremely small security blocks. Petrie's published design targets approximately 40,000 transistors per block~\cite{petrie2025embedded}, which the source characterises as a negligible die area overhead; in our assessment, however, this budget appears tight once constant-time cryptographic implementations, glitch detectors, and random masking are added, based on the area costs documented in the smartcard and payment terminal literature. Experience from smartcards and payment terminals suggests that side-channel resistance consumes area, power, and design effort that these proposals currently abstract away. FlexHEG-style enclosure-based designs (Architecture~B) are less exposed because the defensive budget can be concentrated in one hardened perimeter, but they too inherit a limitation common to confidential-computing deployments more broadly: sophisticated physical attacks are generally considered out of scope for these systems~\cite{ccc2022technical}.

\paragraph{Tier 3: Nation-state adversaries} with access to specialised semiconductor failure-analysis laboratories. These actors can realistically conduct FIB circuit editing, systematic decapping and microprobing, or full reverse engineering campaigns. However, these attacks do not scale: individually editing or probing thousands of chips is economically implausible even for highly capable actors. The more serious nation-state threat is \textit{supply-chain compromise}: inserting a backdoor at design, packaging, or foundry level that propagates across an entire production run.

This tiering shifts the policy implication. Physical tamper resistance does not need to be perfect to be governance-relevant; it needs to make circumvention more expensive than compliance for Tier~1 and Tier~2 actors, and sufficiently detectable for Tier~3 actors through inspection, cross-section analysis, or registry inconsistencies. The closest analogy is not absolute tamper-proofing but the IAEA model of tamper-evident assurance: if a signatory attacks its own chips, the goal is that the attack leaves evidence that can later be detected~\cite{aarne2024secure}.

Petrie et al.\ address tamper resistance in the FlexHEG design through systems that wipe secrets and optionally blow fuses if intrusion is detected~\cite{petrie2025flexible}, but acknowledge that ``early versions may need to compromise on security''~\cite{petrie2025flexible}. The RAND HEM report states that ``more research is needed to ensure that HEMs can remain secure even when in the physical possession of an adversary''~\cite{kulp2024hardware}. The Confidential Computing Consortium identifies side-channel vulnerabilities in existing TEEs, noting that an attacker can ``accurately measure the power usage of the TEE CPU during method execution'' to infer information about protected data~\cite{ccc2022technical}.

\subsection{Privacy, Sovereignty, and Political Feasibility}
\label{sec:political}

Hardware-level governance mechanisms that provide the strongest verification and enforcement capabilities are also the mechanisms that raise the most severe political objections. The tensions are structural, not incidental, and cannot be resolved through better engineering alone.

\paragraph{Privacy.} Compute monitoring (M1--M2), workload classification, and hardware-based logging all involve collecting information about what customers do with their computing resources. Sastry et al.\ identify threats to personal privacy, leakage of sensitive strategic and commercial information, and risks from centralisation and concentration of power~\cite{sastry2024computing}. Heim, Anderljung, and Belfield warn that ``intrusive compute governance measures risk infringing on civil liberties''~\cite{heim2024govern}. Heim et al.\ note that balancing regulatory requirements with customer privacy laws, particularly concerning EU-US data transfer frameworks (GDPR), creates additional legal complexity~\cite{heim2024governing}. Privacy-preserving approaches exist (V1 TEEs, zero-knowledge proofs, V3 FlexHEG's anonymised pings), but each involves trade-offs against verification strength; perfect privacy and perfect verification are fundamentally in tension.

\paragraph{Sovereignty.} Remote disablement (E4), location-based chip locking, and hardware-embedded enforcement concentrate control over a critical infrastructure layer in the hands of a small number of chip-producing states, currently dominated by the United States. This creates asymmetric geopolitical leverage that chip-importing states will resist. Bernabei et al.\ note that compute governance proposals face pushback from countries worried about ``AI protectionism'' and American dominance~\cite{bernabei2024legal}. Petrie et al.\ acknowledge that the original vision of fully automatic on-device compliance checking is ``unlikely to be feasible'' partly because it would be ``perceived as a violation of sovereignty''~\cite{petrie2025flexible}. The IAEA analogy is instructive but imperfect: the IAEA operates with the consent of signatory states and does not possess a unilateral ability to disable nuclear facilities remotely. Governance architectures that distribute enforcement authority across multiple independent jurisdictions, such as Petrie et al.'s proposal to require approval from all or a majority of a defined set of governments before certain training actions proceed~\cite{petrie2025flexible}, represent one approach to addressing the political feasibility problem, though no detailed design has been implemented.

\paragraph{Regulatory capture and abuse.} Centralised control over compute access creates opportunities for regulatory capture by incumbent AI firms, suppression of academic and open-source research, and abuse by authoritarian governments. Sastry et al.\ explicitly warn of these risks, identifying threats from centralisation and concentration of power, including the possibility that corrupt or oppressive policymakers could misuse governance tools such as chip registries~\cite{sastry2024computing}. Heim, Anderljung, and Belfield warn that intrusive compute governance measures risk infringing on civil liberties and that centralising control of compute could entrench existing power asymmetries~\cite{heim2024govern}. Lehdonvirta et al.\ highlight distributional concerns raised by compute-based governance: countries in the ``Compute South'' are less well-positioned to shape AI system development, and those in the ``Compute Desert'' have few prospects for making use of compute governance as a means to influence AI~\cite{lehdonvirta2024compute}.

\subsection{The Concentration Dependency}
\label{sec:concentration}

The entire architecture of hardware-level governance rests on a structural contingency: the extreme concentration of advanced semiconductor manufacturing. As of 2023, cutting-edge logic fabrication (sub-7nm) is dominated by TSMC, with Samsung and Intel as the only other players. This concentration makes export controls, chip registries, and supply chain monitoring feasible. However, this concentration is not guaranteed to persist. China's sustained investment in indigenous semiconductor capability is explicitly aimed at reducing this dependency~\cite{shrivastava2025china, cheung2025geopolitical}. Assessments of China's position relative to the frontier vary: Shrivastava describes China's chip manufacturing as ``generations behind'' in key segments such as EUV lithography~\cite{shrivastava2025china}, while Cheung documents significant recent progress, including domestically produced chips approaching parity with recent Western designs and AI models achieving competitive performance within months of leading US releases~\cite{cheung2025geopolitical}. Belfield acknowledges that ``most assessments are that while there is some diversion, it is relatively small and insignificant, and despite substantial investment, indigeneity is years if not decades away''~\cite{belfield2025domestic}. Yoon et al.\ formalise this dynamic in a game-theoretic model of export controls, finding that excessively stringent regimes can backfire by accelerating indigenous innovation in the rival nation, while overly lenient regimes enable persistent circumvention; the most effective controls are those where enforcement is strategically calibrated and compliance costs for regulated firms remain manageable~\cite{yoon2026export}. But a governance architecture designed around manufacturing concentration that does not account for its potential erosion is building on a foundation it cannot control.

The implication is temporal: hardware-level governance mechanisms have a window of opportunity determined by the continued concentration of advanced chip manufacturing. The mechanisms that require the longest development timelines (M5 on-chip metering, V3 FlexHEGs, E6 hardware-embedded licensing) must be developed and deployed before this window narrows. If they are not, the governance community may find itself with sophisticated technical proposals that can no longer be implemented because the manufacturing leverage required to mandate their adoption has diffused.

\subsection{The Training-Inference Distinction}
\label{sec:training-inference}

Most hardware-level governance mechanisms in Section~\ref{sec:taxonomy} are designed for training-time monitoring and enforcement. This reflects the current regulatory assumption that dangerous capabilities are primarily created during training. However, as inference-time compute scaling becomes more capable and fine-tuning and adaptation become more powerful~\cite{pistillo2025defending, jones2025overcoming}, and as mixture-of-experts architectures raise questions about which compute should count for governance purposes~\cite{hooker2024limitations}, the training-inference distinction may cease to be a reliable governance boundary.

Workload classification (M2) can currently distinguish training from inference signatures~\cite{heim2024governing}, but this depends on their computational signatures remaining distinct. As training and inference converge architecturally, this separability may diminish. V3 FlexHEG verification of ``negative claims'' (proving that certain operations were not performed) via compute accounting~\cite{petrie2025flexible} offers one approach to governing inference-time compute; the source describes this as a supported capability of the architecture, though the broader FlexHEG framework is itself still in an early design stage. O'Keefe and Frazier propose that future AI systems could autonomously perform compliance monitoring tasks such as automated evaluations, transparency reporting, and safety incident detection, but acknowledge that this requires significant further AI capability development~\cite{o2026automated}. Such automated compliance could in principle extend governance beyond training into deployment, but the paper addresses compliance automation broadly rather than inference-time governance specifically. The governance literature has not yet produced a comprehensive framework for inference-time compute governance, and this is a gap that will become increasingly urgent as inference-time scaling matures.


\section{Mapping Mechanisms to Governance Scenarios}
\label{sec:mapping}

The mechanisms in Section~\ref{sec:taxonomy} do not exist in a policy vacuum. Their utility depends on the governance context in which they are deployed: what a domestic regulator needs from hardware-level governance differs substantially from what an international treaty verification body requires. This section maps the taxonomy onto four governance scenarios, drawing on analogies to existing verification regimes where instructive.

\subsection{Governance Scenarios}
\label{sec:scenarios}

We consider four scenarios that span the range of current policy proposals:

\begin{enumerate}
    \item \textbf{Domestic frontier AI regulation.} A single jurisdiction imposes obligations on AI developers and compute providers operating within its borders. Examples include the EU AI Act's compute thresholds for general-purpose AI models~\cite{erben2025training}, the US executive order requiring developers and cloud providers to report large training runs~\cite{whitehouse2023eo}, and the UK AI Safety Institute's voluntary evaluation framework~\cite{ritchie2025turing}. The regulator has legal authority over domestic entities but limited reach beyond its borders.

    \item \textbf{Bilateral agreements.} Two states (typically the US and China, or the US and an allied partner) negotiate specific commitments on AI development, potentially including mutual restraint on certain training scales, export control coordination, or reciprocal verification. The US chip export controls targeting China already function as a de facto constraint on bilateral compute access~\cite{belfield2025domestic}.

    \item \textbf{Multilateral treaty with verification body.} An international agreement, potentially modelled on the Nuclear Non-Proliferation Treaty (NPT) or the Chemical Weapons Convention (CWC), establishes a verification body with inspection authority across signatory states. Belfield proposes this directly as a ``Secure Chips Agreement: a Non-Proliferation Treaty for state-of-the-art AI chips''~\cite{belfield2025domestic}. Ramiah et al.\ design a ``Global Compute Pause Button'' framework along similar lines~\cite{ramiah2025toward}.

    \item \textbf{Industry self-regulation.} AI developers and compute providers voluntarily adopt governance mechanisms, potentially under pressure from customers, investors, or reputational risk, but without legal mandate. Related examples include voluntary commitments from frontier AI companies at the UK-hosted Seoul AI Summit, and the growing role of third-party evaluators and auditors in AI risk management~\cite{caputo2025risk}.
\end{enumerate}

\subsection{Mechanism--Scenario Mapping}
\label{sec:matrix}

Table~\ref{tab:mapping} presents the mapping of mechanisms to governance scenarios, assessed on a three-point scale. A \textit{strong fit} ($++$) indicates that the mechanism is both technically appropriate and institutionally aligned with the scenario: it addresses a core requirement of the governance context and can be implemented within its institutional constraints. \textit{Applicable with caveats} ($+$) indicates that the mechanism is relevant but faces significant limitations in the scenario, such as jurisdictional reach, adversarial robustness, or dependence on voluntary cooperation. \textit{Limited applicability} ($-$) indicates that the mechanism is either technically unsuitable for the scenario or faces structural barriers to adoption within it. These ratings synthesise the feasibility analysis from Section~\ref{sec:taxonomy} with the adversarial considerations from Section~\ref{sec:adversarial} and the institutional requirements of each scenario.

\begin{table}[htbp]
\centering
\caption{Appropriateness of hardware-level governance mechanisms across governance scenarios. $++$ = strong fit; $+$ = applicable with caveats; $-$ = limited applicability or infeasible. Mechanism identifiers correspond to Section~\ref{sec:taxonomy}.}
\label{tab:mapping}
\small
\begin{tabular}{p{4.5cm} c c c c}
\toprule
\textbf{Mechanism} & \textbf{Domestic} & \textbf{Bilateral} & \textbf{Treaty} & \textbf{Self-reg.} \\
\midrule
\multicolumn{5}{l}{\textit{Monitoring}} \\
\addlinespace
M1 Cloud provider metadata & $++$ & $+$ & $-$ & $++$ \\
M2 Workload classification & $++$ & $+$ & $+$ & $+$ \\
M3 KYC for compute & $++$ & $+$ & $+$ & $-$ \\
M4 Power consumption monitoring & $+$ & $+$ & $++$ & $-$ \\
M5 On-chip FLOP metering & $+$ & $+$ & $++$ & $-$ \\
M6 Chip location tracking & $+$ & $++$ & $++$ & $-$ \\
M7 Chip registry / supply chain & $+$ & $+$ & $++$ & $-$ \\
\addlinespace
\multicolumn{5}{l}{\textit{Verification}} \\
\addlinespace
V1 TEE workload attestation & $++$ & $+$ & $+$ & $++$ \\
V2 Proof-of-training & $+$ & $+$ & $++$ & $+$ \\
V3 FlexHEG verifiable claims & $+$ & $+$ & $++$ & $+$ \\
V4 Remote attestation & $+$ & $++$ & $++$ & $-$ \\
V5 Multiparty control & $-$ & $+$ & $++$ & $-$ \\
V6 Physical inspection & $+$ & $+$ & $++$ & $-$ \\
\addlinespace
\multicolumn{5}{l}{\textit{Enforcement}} \\
\addlinespace
E1 Cloud access control & $++$ & $+$ & $-$ & $++$ \\
E2 Networking restrictions & $-$ & $+$ & $+$ & $-$ \\
E3 Hardware off-switches & $-$ & $+$ & $++$ & $-$ \\
E4 Remote disablement & $-$ & $++$ & $+$ & $-$ \\
E5 Export controls & $-$ & $++$ & $+$ & $-$ \\
E6 Compute licensing & $++$ & $+$ & $+$ & $+$ \\
E7 Upstream supply chain & $-$ & $++$ & $++$ & $-$ \\
\bottomrule
\end{tabular}
\end{table}

Several patterns are visible. \textit{Domestic regulation} is best served by mechanisms that leverage existing infrastructure: M1 cloud provider metadata, M3 KYC, M2 workload classification, V1 TEE attestation, and E6 compute licensing. These are either currently deployable or near-term and operate within the regulator's jurisdictional reach. The principal limitation is that domestic mechanisms cannot govern compute accessed in other jurisdictions or on-premises hardware outside the cloud.

\textit{Bilateral agreements}, particularly between the US and China, require mechanisms that can be verified without full trust: M6 chip location tracking, V4 remote attestation, E5 export controls, and E7 upstream supply chain controls. The current US export control regime already implements a unilateral version of this scenario. E4 remote disablement scores highly for bilateral enforcement because it provides ongoing leverage beyond the point of sale, but as discussed in Section~\ref{sec:political}, its political feasibility depends on whether the authority to disable can be structured as bilateral rather than unilateral.

The \textit{multilateral treaty} scenario has the most demanding requirements and is best served by the mechanisms that are, correspondingly, the least mature. M5 on-chip FLOP metering, V2 proof-of-training, V3 FlexHEG verification, V6 physical inspection, E3 hardware off-switches, and E7 supply chain controls form the core of a treaty verification architecture. V5 multiparty cryptographic control, the most speculative mechanism in the taxonomy, finds its primary use case here, implementing the ``two-person rule'' at the hardware level to prevent unilateral action.

\textit{Industry self-regulation} is the most limited scenario, restricted to mechanisms that require no legal mandate and can be adopted voluntarily by individual firms: M1 cloud provider metadata, V1 TEE attestation, E6 compute licensing (as internal policy), and E1 cloud access control. Mechanisms requiring hardware modification or cross-firm coordination are effectively unavailable under self-regulation.

\subsection{Lessons from Existing Verification Regimes}
\label{sec:analogies}

The governance scenarios described above, particularly the multilateral treaty scenario, are not without precedent. Several existing international verification regimes offer structural lessons for hardware-level AI governance, though each analogy has significant limitations.

\paragraph{Nuclear non-proliferation (NPT/IAEA).}
The NPT/IAEA regime is a widely invoked reference point in the compute governance literature~\cite{belfield2025domestic, ramiah2025toward, baker2023nuclear, vermeer2024historical, ho2023international}. The structural parallels are genuine: both nuclear material and advanced AI chips are dual-use, require substantial capital investment, are quantifiable, and are produced through a concentrated supply chain that provides natural chokepoints for governance~\cite{sastry2024computing}. IAEA verification combines material accountancy (tracking declared nuclear materials through facilities), on-site inspections, remote surveillance, environmental sampling, and satellite imagery~\cite{baker2023nuclear, ramiah2025toward}. Baker argues that ``with certain preparations, the foreseeable challenges of verification would be reduced to levels that were successfully managed in nuclear arms control''~\cite{baker2023nuclear}. In Table~\ref{tab:mapping} terms, the IAEA model most directly supports the ``Treaty'' column: M7 chip registries (analogous to material accountancy), V6 physical inspection, M4 power monitoring (analogous to environmental sampling), and E7 upstream supply chain controls (analogous to enrichment restrictions) all have high ratings in the treaty scenario precisely because they mirror IAEA practice.

The analogy's limitations are equally important. First, nuclear material is radioactive and physically detectable at borders; AI chips are not, making diversion harder to detect through passive means~\cite{sastry2024computing}. Second, the IAEA operates with the consent of signatory states and does not possess unilateral enforcement authority; any ``IAEA for AI'' would face similar constraints. Third, the nuclear non-proliferation regime required decades to build and still has significant gaps; Belfield acknowledges that building analogous AI governance institutions may similarly take ``years or decades,'' yet the pace of AI capability growth creates pressure for faster institutional development than the nuclear precedent required~\cite{belfield2025domestic}. Fourth, model weights are digital information that, once accessed, can be copied and distributed without physical constraints analogous to those governing fissile material; Nevo et al.\ document the security challenges of protecting frontier model weights from theft and unauthorised access, though they note that the large size of frontier models (reaching terabytes) may make unauthorised duplication somewhat easier to detect than for smaller digital assets~\cite{nevo2024securing}.

\paragraph{Chemical weapons (CWC/OPCW).}
The CWC/OPCW regime offers additional structural lessons for AI governance~\cite{ramiah2025toward, wasil2024verification, scholefield2025international}. The OPCW manages dual-use chemicals through a schedules system (classifying substances by risk level), on-site inspections, and challenge inspections~\cite{wasil2024verification}; Ramiah et al.\ note that its focus on precursor control provides a strong analogy for compute governance, and that the OPCW's inspection capabilities offer an important precedent for hardware-level verification~\cite{ramiah2025toward}.

\paragraph{Financial regulation (KYC/AML).}
Heim et al.\ draw a structural analogy between compute providers and financial institutions, both serving as intermediaries through which regulated activity must pass~\cite{heim2024governing}. The Financial Action Task Force (FATF) model of international coordination, where member states agree to implement common anti-money-laundering standards and face peer review for compliance, is suggested by Heim et al.\ as a possible model for international compute oversight coordination, facilitating information sharing and helping align regulatory approaches without requiring a formal treaty~\cite{heim2024governing}. This analogy maps directly onto the ``Domestic'' column in Table~\ref{tab:mapping}: M3 KYC and M1 cloud provider metadata both rate as strong fits for domestic regulation, and the FATF model suggests a pathway for international harmonisation of these domestic measures without requiring a formal treaty. Heim et al.\ note that the financial practice of splitting transactions to evade automated reporting thresholds is directly analogous to how actors might split training runs across providers to evade compute governance~\cite{heim2024governing}. The limitation is that financial regulation operates through institutional intermediaries (banks); AI governance must also account for on-premises compute that does not pass through any intermediary.

\paragraph{Implications for institutional design.}
Vermeer systematically compares nuclear technology, the Internet, encryption, and genetic engineering as governance analogues, concluding that no single model is a perfect fit for AI but each offers important lessons depending on the risk profile that materialises~\cite{vermeer2024historical}. Ho et al.\ propose four institutional models for international AI governance, each with precedents in existing organisations: a scientific assessment body (analogous to the IPCC), a standards and monitoring organisation (analogous to the IAEA or FATF), a development and access initiative (analogous to Gavi or the IAEA's nuclear fuel bank), and a safety research project (analogous to CERN)~\cite{ho2023international}.

\begin{figure}[htbp]
	\centering
	\includegraphics[width=\textwidth]{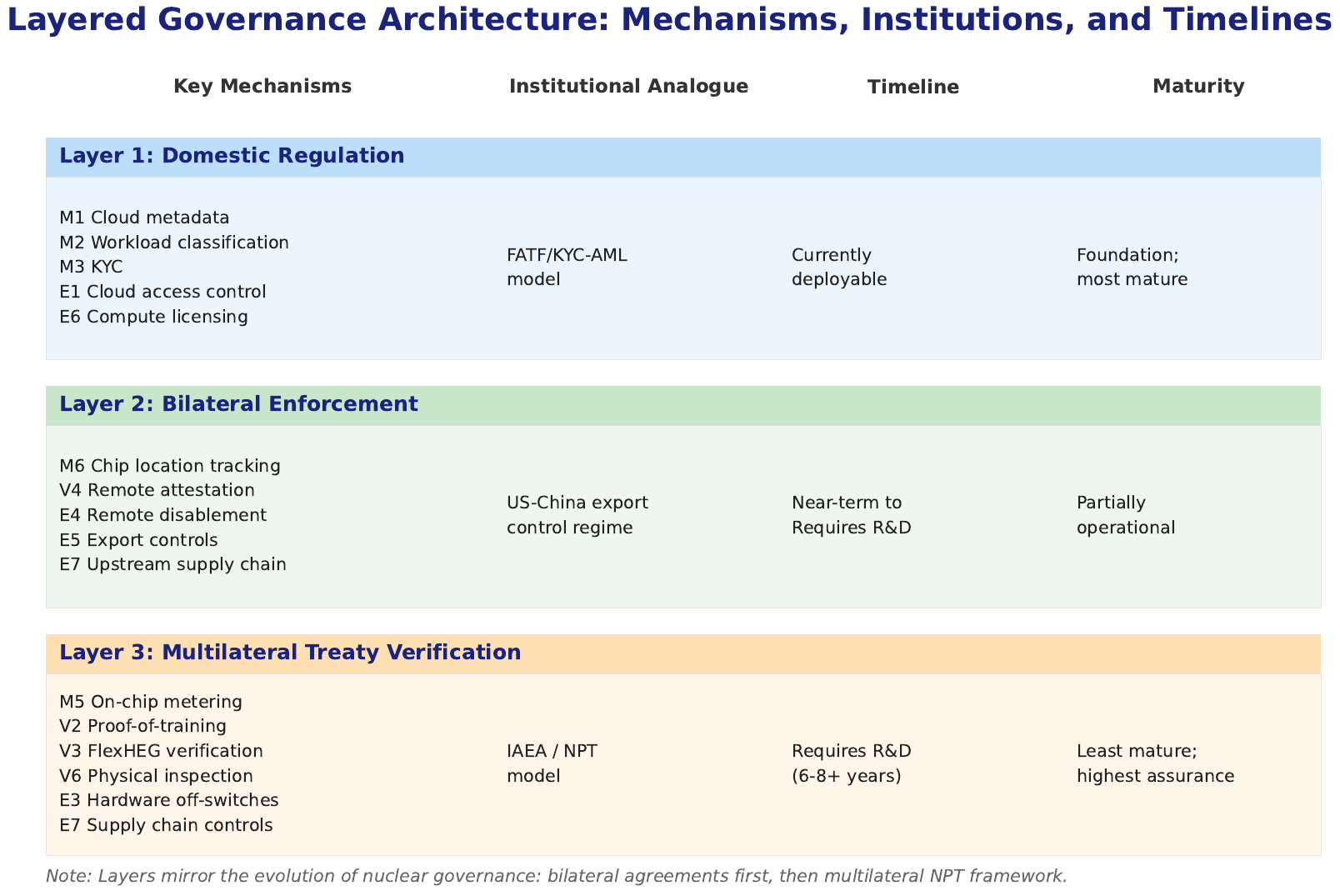}
	\caption{Layered governance architecture mapping mechanisms to institutional models. Layer~1 (domestic regulation) builds on currently deployable mechanisms following the FATF model. Layer~2 (bilateral enforcement) extends governance through export controls and hardware tracking. Layer~3 (multilateral verification) requires the most technically demanding mechanisms, following the IAEA model. Layers are designed to be built sequentially, with each providing a foundation for the next.}
	\label{fig:layered-governance}
\end{figure}

The mapping in Table~\ref{tab:mapping} suggests that the institutional architecture most likely to succeed will be layered rather than monolithic, as illustrated in Figure~\ref{fig:layered-governance}.

The first layer, \textit{domestic regulation}, uses currently deployable mechanisms (M1, M2, M3, E1, E6) as the foundation, following the FATF model of harmonised national standards. The second layer, \textit{bilateral enforcement}, uses export controls (E5), hardware tracking (M6), and upstream supply chain controls (E7), extending governance beyond individual jurisdictions. The third layer, \textit{multilateral verification}, uses the more technically demanding mechanisms (V2, V3, V6, M5, E3) that will take longer to build but provide the highest assurance, following the IAEA model of formal inspections and material accountancy. This layered approach mirrors the evolution of nuclear governance, which began with bilateral US-Soviet agreements before expanding into the multilateral NPT framework, and recognises that waiting for the ideal multilateral treaty before implementing any governance would leave a dangerous gap during the period of most rapid AI capability growth.


\section{Discussion and Implications}
\label{sec:discussion}

\subsection{The Readiness Gap and the Closing Window}

The central finding of this paper is a structural mismatch between what governance frameworks demand and what technology can currently deliver, compounded by a temporal constraint that makes the mismatch urgent rather than merely inconvenient.

Domestic regulation can function with currently deployable mechanisms: M1 cloud provider metadata, M3 KYC requirements, M2 workload classification, and E1 administrative access controls. These are sufficient for the reporting and licensing regimes established by the EU AI Act and (prior to revocation) US Executive Order~14110. But the governance scenarios with the highest stakes, bilateral enforcement between strategic competitors and multilateral treaty verification, require mechanisms that remain in early-stage research: M5 on-chip compute metering, V2 cryptographic proof-of-training protocols, V3 FlexHEG-based verification, and E3/E6 hardware-embedded enforcement.

This readiness gap has a temporal dimension that the governance community has not adequately reckoned with. The feasibility of hardware-level governance depends on semiconductor manufacturing concentration, which provides the leverage to mandate mechanism adoption in new chip designs. Three trends are simultaneously narrowing this window: distributed training methods are maturing, algorithmic efficiency is reducing the compute required for frontier capabilities, and chip manufacturing capacity is beginning to diffuse beyond its current concentration in a small number of allied democracies (Section~\ref{sec:concentration}). Aarne et al.\ estimate 18~months to 4~years of development time for on-chip mechanisms, followed by a further 4~years for sufficiently widespread deployment~\cite{aarne2024secure}. A total timeline of 6 to 8+ years for the most critical mechanisms is not obviously inconsistent with the window, but it leaves essentially no margin for delay. If development does not begin in the near term, the mechanisms that governance frameworks most need may arrive after the structural conditions that make them implementable have eroded.

\begin{figure}[htbp]
	\centering
	\includegraphics[width=\textwidth]{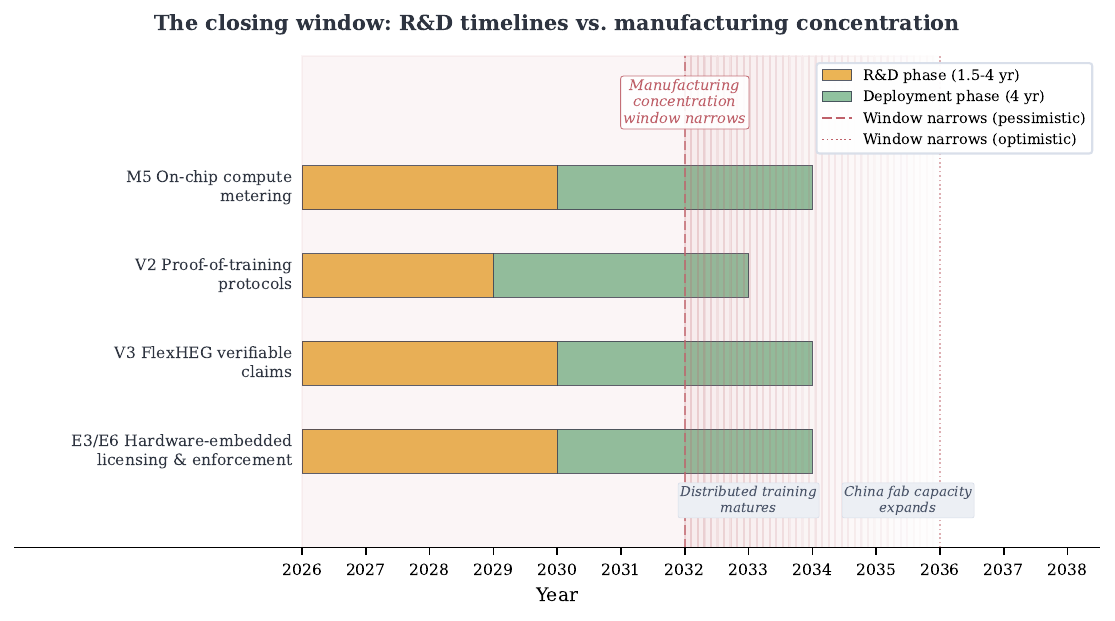}
	\caption{R\&D and deployment timelines for the four highest-priority mechanisms (from Section~\ref{sec:feasibility-summary}) against the estimated window of semiconductor manufacturing concentration. R\&D phase estimates (1.5--4 years) are drawn from Aarne et al.~\cite{aarne2024secure}; a further 4-year deployment phase reflects their estimate for sufficiently widespread adoption. The pessimistic window boundary (dashed) represents a scenario where distributed training maturation and indigenous fab capacity building in restricted jurisdictions erode manufacturing leverage by approximately 2032; the optimistic boundary (dotted) extends this to approximately 2036. In either scenario, the margin between mechanism readiness and window closure is narrow, underscoring the urgency of near-term R\&D investment.}
	\label{fig:temporal-window}
\end{figure}

\subsection{Implications for Research}

Four lines of technical research emerge as priorities from the feasibility analysis:

First, \textbf{on-chip compute metering} (M5) requires progression from conceptual designs to functional prototypes. The FlexHEG architecture~\cite{petrie2025flexible} and the embedded off-switch design~\cite{petrie2025embedded} provide starting points, but neither has been fabricated or tested against physical attacks. The RAND assessment that ``more research is needed to ensure that HEMs can remain secure even when in the physical possession of an adversary''~\cite{kulp2024hardware} underscores a major unresolved constraint.

Second, \textbf{proof-of-training protocols} (V2) must advance from theoretical proposals to practical implementations. Jia et al.'s Proof-of-Learning~\cite{jia2021proof} and Shavit's Proof-of-Training-Transcript~\cite{shavit2023does} provide initial conceptual frameworks, but current PoL schemes are heuristic-based and not yet provably secure~\cite{shavit2023does}, and production-grade systems that can operate at the scale of frontier training runs without prohibitive overhead do not yet exist. South et al.'s work on zkSNARK-based verifiable evaluations~\cite{south2024verifiable} suggests a pathway for specific verification tasks, while general-purpose training verification lies outside the scope of their paper, which focuses on model evaluation rather than training.

Third, \textbf{inference-time governance mechanisms} are almost entirely absent from the current literature. As inference-time compute scaling, mixture-of-experts architectures, and post-training enhancement techniques blur the boundary between training and deployment (Section~\ref{sec:training-inference}), governance mechanisms designed exclusively for training-time monitoring will become increasingly insufficient~\cite{hooker2024limitations, pistillo2025defending}. Extending V3 FlexHEG verification and V2 proof-of-training concepts to the inference domain is a necessary but largely unexplored research direction.

Fourth, \textbf{governance architecture design} for hardware-embedded enforcement, particularly E4 remote disablement and V5 multiparty control, must address the sovereignty and legitimacy concerns identified in Section~\ref{sec:political}. Technical designs that concentrate enforcement authority in a single state or entity will be rejected by the international community regardless of their technical merit. Distributed governance architectures, potentially building on Petrie et al.'s proposal for training licences that require approval from a majority of a defined set of governments~\cite{petrie2025flexible}, require both cryptographic protocol design and institutional design, which are often discussed in separate strands of the literature.

\subsection{Implications for Policy}

For policymakers, the taxonomy and feasibility classification offer three actionable insights. First, regulations that depend on mechanisms rated ``requires R\&D'' or ``speculative'' should not be enacted as if those mechanisms already exist; doing so creates unenforceable mandates that erode regulatory credibility. The appropriate policy action is to fund the R\&D, not to legislate the outcome prematurely. Second, the layered governance approach suggested by the mechanism-to-scenario mapping in Section~\ref{sec:mapping}, where domestic regulation builds on currently deployable mechanisms while international frameworks are constructed in parallel using more technically demanding ones, provides a pragmatic pathway that does not require waiting for all mechanisms to mature before any governance is implemented. Third, international coordination on chip registry standards (M7), KYC requirements for compute providers (M3), and hardware security specifications should begin now, even before the more ambitious mechanisms are ready, to establish the institutional infrastructure that those mechanisms will eventually require.

\subsection{Limitations}

This paper has several limitations that should inform how its conclusions are interpreted. First, the feasibility assessments are based on published literature and publicly available technical specifications; classified or proprietary work within chip manufacturers or intelligence agencies may have advanced certain mechanisms further than the open literature indicates. Second, the taxonomy focuses on hardware for training and inference of large neural networks; specialised hardware for other AI paradigms (neuromorphic computing, quantum computing, optical computing) is not systematically addressed. Kashif et al.\ argue that the governance challenges of neuromorphic computing are already urgent as these systems move from laboratory settings into real-world applications~\cite{kashif2026governance}; the governance implications of quantum and optical computing substrates remain less explored and are not addressed here. Third, the governance-scenario mapping in Section~\ref{sec:mapping} assesses mechanism appropriateness but does not model the political economy of adoption: which actors would support or oppose each mechanism, and under what conditions agreement might be reached. Fourth, the paper does not address the environmental and economic costs of hardware-level governance mechanisms (die area overhead, power consumption, manufacturing complexity), which will factor into industry willingness to adopt them. Fifth, the feasibility ratings are derived from the author's assessment of the reviewed literature rather than from formal expert elicitation or independent technology readiness review; where this paper reports specific engineering parameters (transistor counts, gate counts, development timelines), these are drawn from the cited sources and should be treated as reported estimates rather than independently verified figures.

\subsection{Conclusion}

Hardware-level governance of AI compute is technically feasible in principle but unevenly mature in practice. A substantial body of mechanisms exists at the conceptual and early-design stage; a much smaller set is currently deployable. The gap between these two categories defines the near-term research agenda for the field. Closing this gap is not merely an academic exercise: the political window during which semiconductor manufacturing concentration makes hardware-level governance implementable will not remain open indefinitely. The mechanisms described in this paper, if developed and deployed within that window, could provide the technical substrate for verifiable international agreements on AI development. If they are not, the governance community will be left with policy instruments that lack the technical means of verification and enforcement, a condition that, as the history of arms control demonstrates, is a recipe for regime failure.

\bibliographystyle{unsrt}
\bibliography{AI_Gov_final_reference_list}

\end{document}